\documentclass[conference]{IEEEtran}
\IEEEoverridecommandlockouts
\usepackage{cite}
\usepackage{amssymb,amsmath,amstext,amsfonts}
\usepackage{graphicx}
\usepackage{textcomp}
\usepackage{xcolor}
\def\BibTeX{{\rm B\kern-.05em{\sc i\kern-.025em b}\kern-.08em
    T\kern-.1667em\lower.7ex\hbox{E}\kern-.125emX}}

\usepackage{cite}
\usepackage{url}

\usepackage{mathtools}
\usepackage{bm}  
\usepackage{mathrsfs} 
\usepackage{siunitx}

\usepackage{url}
\usepackage{filecontents}
\usepackage{stmaryrd}

\usepackage{graphicx}
\usepackage{epstopdf}
\usepackage[colorinlistoftodos]{todonotes}
\usepackage{easyReview}

\usepackage{amsmath,amssymb}

\usepackage{siunitx}
\DeclareSIUnit{\mmmrad}{\milli\meter\textrm{-}\milli\radian}

\graphicspath{%
	{Figures/}
	{Figures/TiKz/}
	{Figures/Ipe/}
	{Figures/XFig/}
	{Figures/Inkscape/}
	{Figures/Matlab/}
	{Figures/Scanned/}
	{Figures/Saved/}
}

\DeclareMathOperator*{\argmin}{arg\,min}

\newsavebox{\myparbox}
\newlength{\myparboxwidth}

\providecommand{\norm}[1]{\lVert#1\rVert}

\newcommand\Angle[1]{\setbox0=\hbox{$\mskip 7mu minus 4mu#1$}%
  \raise.21ex\hbox{$/$}\hskip-0.95ex\underline{\raise\dp0\hbox{\box0}}}

\ExplSyntaxOn

\NewDocumentCommand \vect { s o m }
 {
  \IfBooleanTF {#1}
   { \vectaux*{#3} }
   { \IfValueTF {#2} { \vectaux[#2]{#3} } { \vectaux{#3} } }
 }

\DeclarePairedDelimiterX \vectaux [1] {\lbrack} {\rbrack}
 { \, \dbacc_vect:n { #1 } \, }

\cs_new_protected:Npn \dbacc_vect:n #1
 {
  \seq_set_split:Nnn \l_tmpa_seq { , } { #1 }
  \seq_use:Nn \l_tmpa_seq { \enspace }
 }
\ExplSyntaxOff

\usepackage{cite}
\usepackage{amsmath,amssymb,amsfonts,amstext}
\usepackage{graphicx}
\usepackage{textcomp}
\usepackage{xcolor}
\def\BibTeX{{\rm B\kern-.05em{\sc i\kern-.025em b}\kern-.08em
		T\kern-.1667em\lower.7ex\hbox{E}\kern-.125emX}}

\usepackage{multicol}

\usepackage{url}
\usepackage{booktabs}

\usepackage{mathtools}
\usepackage{bm}  
\usepackage{siunitx}

\usepackage{filecontents}
\usepackage{stmaryrd}

\usepackage{tabu}
\usepackage{environ}

\usepackage{enumerate}
\usepackage{ifthen}

\usepackage{multirow}

\usepackage{physics}
\usepackage{siunitx}
\usepackage{nomencl} 
\makenomenclature

\usepackage{algorithm,algpseudocode,float}
\usepackage{setspace} 

\algnewcommand{\algorithmicand}{\textbf{ and }}
\algnewcommand{\algorithmicor}{\textbf{ or }}
\algnewcommand{\AlgAnd}{\algorithmicand}
\algnewcommand{\AlgOr}{\algorithmicor}

\usepackage[firstpageonly=true]{draftwatermark}				

\usepackage{hyperref}
\hypersetup{
    colorlinks=true,
    linkcolor=blue,
    filecolor=magenta,      
    urlcolor=cyan,
    bookmarks=true,
}
\usepackage[noabbrev]{cleveref}  

\begin{document}

\title{\mbox{}\\Trajectory Tracking of Underactuated Sea Vessels With Uncertain Dynamics: An Integral Reinforcement Learning Approach
}

\author{Mohammed Abouheaf, Wail Gueaieb,  Md Suruz Miah, and Davide Spinello 
	\thanks{Mohammed Abouheaf and Wail Gueaieb are with School of Electrical Engineering and Computer Science, University of Ottawa, Ottawa, Ontario, Canada. e-mail: \{mabouhea,wgueaieb\}@uOttawa.ca}
			\thanks{Md Suruz Miah is with Department of Electrical and Computer Engineering, Bradley University, Peoria, Illinois, USA. e-mail: smiah@bradley.edu}
		\thanks{Davide Spinello is with Department of Mechanical Engineering, University of Ottawa, Ottawa, Ontario, Canada. e-mail: dspinell@uOttawa.ca}
}
\maketitle

\DraftwatermarkOptions{%
angle=0,
hpos=0.5\paperwidth,
vpos=0.97\paperheight,
fontsize=0.012\paperwidth,
color={[gray]{0.2}},
text={
  \newcommand{\thispaperdoi}{10.1109/SMC42975.2020.9283399}
  \parbox{0.99\textwidth}{This is the postscript version of the published paper. (doi: \href{http://dx.doi.org/\thispaperdoi}{\thispaperdoi})\\
    \copyright 2020 IEEE.  Personal use of this material is permitted.  Permission from IEEE must be obtained for all other uses, in any current or future media, including reprinting/republishing this material for advertising or promotional purposes, creating new collective works, for resale or redistribution to servers or lists, or reuse of any copyrighted component of this work in other works.}},
}

\begin{abstract}
Underactuated systems like sea vessels have degrees of motion that are insufficiently matched by a set of independent actuation forces. In addition, the underlying trajectory-tracking control problems grow in complexity in order to decide the optimal rudder and thrust control signals. This enforces several difficult-to-solve constraints that are associated with the error dynamical equations using classical optimal tracking and adaptive control approaches. An online machine learning mechanism based on integral reinforcement learning is proposed to find a solution for a class of nonlinear tracking problems with partial prior knowledge of the system dynamics. The actuation forces are decided using innovative forms of temporal difference equations relevant to the vessel's surge and angular velocities. The solution is implemented using an online value iteration process which is realized by employing means of the adaptive critics and gradient descent approaches. The adaptive learning mechanism exhibited well-functioning and interactive features in react to different desired reference-tracking scenarios.
\end{abstract}


\section{Introduction}
The trajectory-tracking problem is a sub-class of the optimal tracking problems where error dynamical equations are derived to solve this category of problems. A considerable number of solution methods are either offline or rely on complicated adaptive structures. The complexity escalates when it is desired to control underactuated high order systems~\cite{Ding2011,Li18,Ning2019}. This work proposes a data-driven machine learning approach that fuses measurements into control strategies for underactuated sea vessels. This approach makes use of the tracking error signals of the orientation of the vessel to compensate for the thrust and rudder forces without actually solving any error dynamical equations or solving complex adaptive control laws. This work combines ideas from Reinforcement Leaning (RL) and optimal control theory to propose reference-tracking control mechanism for a class of underactuated mechanical systems.  

Numerous control approaches, such as backstepping, sliding mode, nonlinear adaptive controllers, among others, have been proposed in the literature to control the trajectories of unmanned  vessels~\cite{Ding2011,Li18,Ning2019}. Among them, approximate dynamic programming approaches are used to tackle the action-state curses of dimensionality associated with the dynamic programming methods. They are employed to solve the optimal control problems using different forms of Bellman equations to optimize the underlying cost functions~\cite{Werbos1974,Bert_1995}. The solution of the Hamilton-Jacobi-Bellman (HJB) of a dynamical system, not only provides an optimal control solution but also forms a temporal difference optimization setup that is employed by machine learning environments~\cite{Bryson1996,Lewis_2012,AbouheafCTT2015}. The optimal solution is found by applying Bellman optimality conditions to derive the optimal strategy. Bellman equations may be arranged into various temporal difference forms that can be realized using reinforcement learning approaches~\cite{AbouheafCH2014}. 

The RL mechanisms allow dynamical systems to select their control strategies in a dynamic learning environment to transit to better states that maximize the sum of cumulative rewards~\cite{Sutton_1998,AbouheafIRIS17,AbouheafCDC14}. Integral Reinforcement Learning (IRL) approaches are employed to solve the differential graphical games and adaptive control problems for linear and nonlinear systems in~\cite{Vrabie2009,AbouheafECC14,Park15,AbouheafIET19}. Policy and value iteration methods provide different realizations for the IRL solutions. These are used to approximate the best strategies-to-follow, by minimizing a performance index of the states or dynamic positions~\cite{AbouheafCTT2015}. Policy iteration method necessitates an initial admissible policy and may employ least-square approximations to arrive at certain solutions~\cite{Abouheapolicy2017}. Value iteration provides a non-decreasing sequence of solving value functions bounded above by the optimal solution~\cite{Abouheaf2014}. While in policy iteration, the solving value functions are non-increasing and bounded below by the value at the optimal strategy. The adaptive critics neural network tools are used to approximate the strategy and the associated value using actor and critic structures~\cite{Bertsekas1996,Busoniu2008,Cui17,AbouheafTIM20}. 

The main contributions can be explained as follows; First, a machine learning process is presented to design a reference-tracking mechanism for an underactuated sea vessel. This approach does not use any error dynamical equations or need to employ complicated adaptive control laws. Second, it makes benefit of the online measurements, or simply the tracking errors, to decide the control strategies. Finally, a value iteration process is developed to update the rudder and thrust control strategies using means of adaptive critics.

The remaining of the paper is structured as follows. Section~\ref{sec:shipModelProblemFormulation} explains the dynamics of an underactuated sea vessel followed by laying out the mathematical framework of the trajectory-tracking problem. Sections~\ref{IRL_Bellman}~and~\ref{actor-critic} present the integral reinforcement learning solution and the associated adaptive critics implementation. The analysis of the simulation results and final concluding remarks are introduced in Sections~\ref{sim}~and~\ref{conc}, respectively.

\section{Underactuated Sea Vessel: Problem Setup}
\label{sec:shipModelProblemFormulation}
\begin{figure}[h]
  \centering
  \includegraphics[width=1\linewidth]{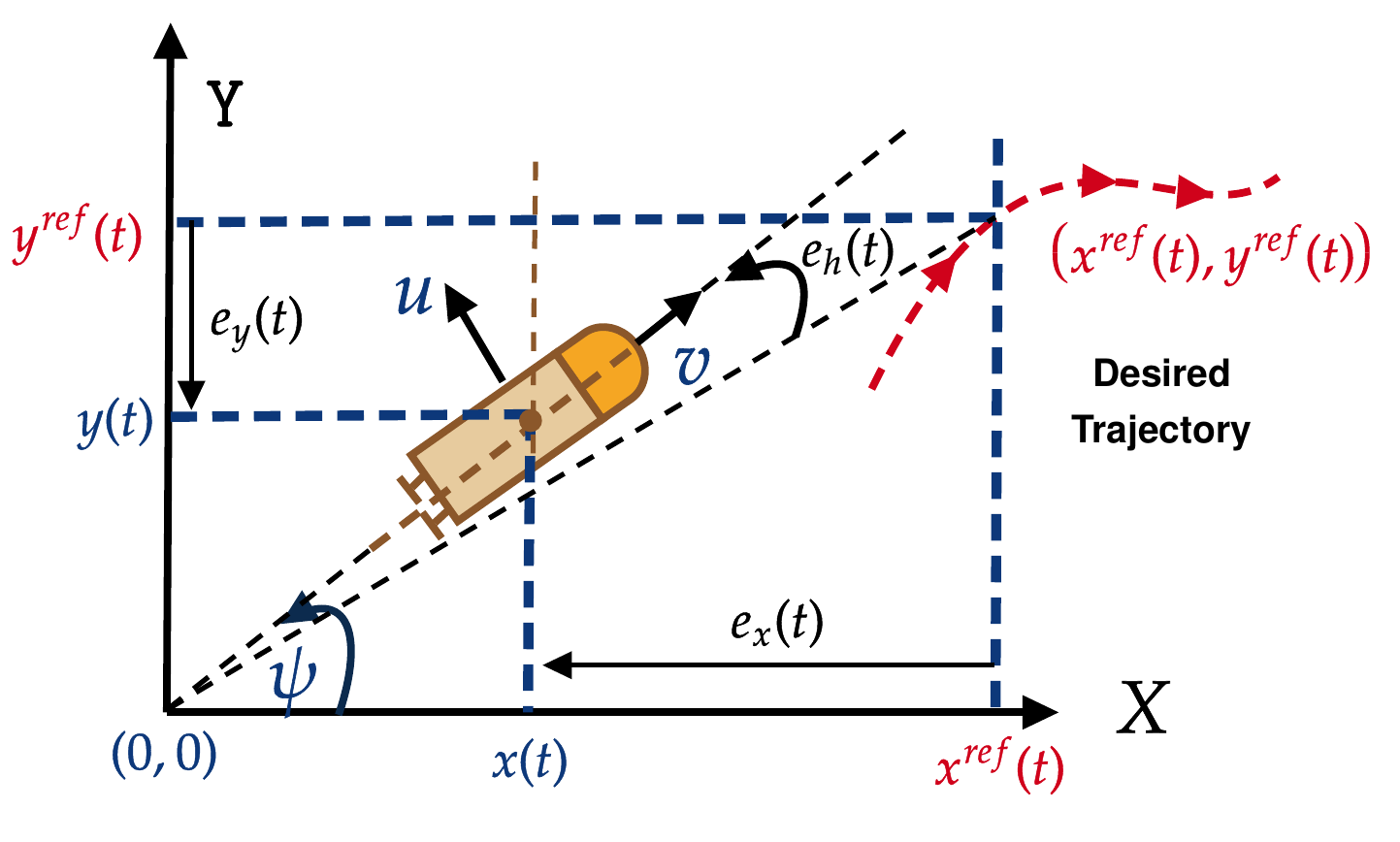}
  \vspace{-25pt}
  \caption{Kinematics of an underactuated sea vessel in following a reference (desired) trajectory $(x^{\mathrm{ref}}(t),y^{\mathrm{ref}}(t)),~t\ge 0.$}
  \label{fig:ship}
\end{figure}

%
The dynamics of an underactuated ship are defined by~\cite{Ding2011}
\begin{subequations}
  \begin{align}
    \dot x &=\, v \, \cos(\psi)-u \, \sin(\psi)\\ 
    \dot y &=\, v \, \sin(\psi)+u  \, \cos(\psi)\\
    \dot \psi &=\,  r \label{eq:dynamics:psi-dot} \\ 
    \dot v &=\,  \frac{m_{22}}{m_{11}}\,u\,r-\frac{d_{11}}{m_{11}}\,v+\frac{1}{m_{11}}\,c_F
             \label{1d}
    \\
    \dot u &=\,  -\frac{m_{11}}{m_{22}}\,v\,r-\frac{d_{22}}{m_{22}}\,u\\ 
    \dot r &=\,  \frac{m_{11}-m_{22}}{m_{33}} uv-\frac{d_{33}}{m_{33}}\,r+\frac{1}{m_{33}}\, c_R,
             \label{1f}
  \end{align}%
  \label{sysdyn}
\end{subequations}\\
where $(x,y)$ and $\psi$ represent the vessel's position and orientation; $v,$  $r,$ and $u$ denote the surge velocity, yaw angular velocity, and sway velocity, respectively; $m_{11},m_{22},$ and $m_{33},$ represent inertia masses;  $d_{11},d_{22},$ and $d_{33},$ represent drag coefficients; $c_F$ and $c_R$ are the thrust and rudder control forces, respectively. Fig.~\ref{fig:ship} depicts the kinematic parameters of an underactuated vessel with respect to the world coordinate system along with a desired (reference) Cartesian trajectory $(x^{\text{ref}}(t),y^{\text{ref}}(t)),~t\ge 0,$ generated by an independent command generator of the reference mission. 

The objective is to steer the vessel towards the desired reference trajectory so that the tracking errors $e_x(t)=x(t)-x^{\text{ref}}(t)$ and $e_y(t)=y(t)-y^{\text{ref}}(t)$ converge to zero as $t \to \infty$. This task can be formulated as an optimal control problem aiming at calculating the optimal thrust and rudder control forces, $c^*_F$ and $c^*_R$, respectively. Adopting such formulation would lead to nonlinear control laws. To simplify the problem, and so the associated machine learning technique, we will now introduce the concepts of tangential and bearing corrections. 

\subsection{Linear Velocity Optimization: Tangential Corrections}

Define the velocity scalar adjustments along the $x$ and $y$ directions as $c_x(t)=\boldsymbol{K}_x(t) \boldsymbol{E}_x(t)$ and $c_y(t)=\boldsymbol{K}_y(t) \boldsymbol{E}_y(t)$, respectively, with
$\boldsymbol{E}_x(t)=\vect{e_x(t) , e_x(t-\Delta) , e_x(t-2\Delta)}^T \in \mathbb{R}^{3\times 1}$
and
$\boldsymbol{E}_y(t) = \vect{e_y(t) , e_y(t-\Delta) , e_y(t-2\Delta)}^T \in \mathbb{R}^{3\times 1}$,
$\boldsymbol{K}_x(t) \in \mathbb{R}^{1\times 3}$ and $\boldsymbol{K}_y(t) \in \mathbb{R}^{1\times 3}$ are sub-control feedback gains to be later determined using a machine learning process, and $\Delta$ is a sampling time.
The tangential adjustment in the surge velocity $v$ can be written as $c_v(t)=\sqrt{c^2_x(t)+c^2_y(t)}/\Delta-v(t-\Delta)$. This form discounts the old velocity instances and adds a velocity adjustment.  The objective of such formulation is to let $c_v(t) \rightarrow 0$ (or simply $\dot v \to 0$) as $t\rightarrow \infty$ so that the surge velocity dynamics have the form $\dot v=c_v$. Equating to~\eqref{1d}, we get the thrust force $c_F$ as
\begin{align}
\label{eq:cF}
c_F=m_{11}\,c_v-m_{22}\,u\,r+d_{11}\,v.
\end{align}
As such, we can design a controller to minimize objective cost functions $U_x$ and $U_y$  in the $x$ and $y$ directions, respectively, where
$\displaystyle 
	U_i\left(\boldsymbol{E}_i(t),{c}_i(t)\right)=\frac{1}{2}\left(\boldsymbol{E}_i^T(t) \boldsymbol{Q}_i \boldsymbol{E}_i(t)+R_i[c_i(t)]^2\right),
$
$i\in\{x,y\}$, for some design parameters $\boldsymbol{Q}_i > \boldsymbol{0}~\in \mathbb{R}^{3 \times 3}$ and ${R}_i > 0 \in \mathbb{R}$.
When referring to a matrix, the notation ``$> \boldsymbol{0}$'' refers to a positive definite matrix.
The performance index associated to the long-run cost is defined by
\begin{align}
  \label{Jxy}  
  J_i&=\int_{0}^{\infty}U_i\left(\boldsymbol{E}_i(\zeta),{c}_i(\zeta)\right)\, d\zeta.
\end{align}

\subsection{Angular Velocity Optimization: Bearing Correction}
The tracking error corresponding to the vessel's orientation is
$e_h(t) = \Angle{x(t) + j y(t)} - \Angle{x^{\text{ref}}(t) + j y^{\text{ref}}(t)}$, where $\Angle{\cdot}$ denotes the phase of the complex quantity $(\cdot)$.
Define the control law $c_h(t)=\tilde c_h(t) + \Angle{c_x + j c_y}$ with bearing adjustment $\tilde c_h(t)=\boldsymbol{K}_h(t) \boldsymbol{E}_h(t)$, where
$\boldsymbol{E}_h(t) = \vect{e_h(t) , e_h(t-\Delta) , e_h(t-2\Delta) }^T \in \mathbb{R}^{3 \times 1}$
and the gain $\boldsymbol{K}_h(t)  \in \mathbb{R}^{1 \times 3}$ is to be determined later.
This step takes into consideration the orientation error and the bearing corrections as per the surge velocity tangential adjustments $c_x$ and $c_y$. It enables the adoption of a heading control law $\psi(t)=\Angle{x(t) + j y(t)} +c_h(t) \, \in \left[\frac{-\pi}{2},\frac{\pi}{2}\right] \si{\radian}$.
Using \eqref{eq:dynamics:psi-dot} and~\eqref{1f} yields
\begin{equation}
\label{eq:angular-acceleration}
\dot r= \ddot \psi= \frac{m_{11}-m_{22}}{m_{33}} \, u \, v-\frac{d_{33}}{m_{33}}\,r+\frac{1}{m_{33}}\, c_R.
\end{equation}
Applying Euler's approximation, the angular acceleration $\ddot \psi$ is estimated by
$\displaystyle \ddot \psi \approx {\left(\psi(t)-2\psi(t-\Delta)+\psi(t-2\Delta)\right)}/{\Delta^2}.$
Equating to~\eqref{eq:angular-acceleration},
$\displaystyle 
	{\left(\psi(t)-2\psi(t-\Delta)+\psi(t-2\Delta)\right)}/{\Delta^2}= \frac{m_{11}-m_{22}}{m_{33}} \, u \, v-\frac{d_{33}}{m_{33}}\,r+\frac{1}{m_{33}}\, c_R.
$
Therefore, the rudder control force can be written as
\begin{eqnarray}
\nonumber
c_R&=&\frac{\psi(t)-2\psi(t-\Delta)+\psi(t-2\Delta)}{\Delta^2/m_{33}} \\&-& \left(m_{11}-m_{22}\right) \,u\, v+d_{33}\, r. 
\label{eq:cR} 
\end{eqnarray}
The final optimization goal is then to minimize the following objective function associated to the bearing adjustment:
\begin{eqnarray*}
	U_h\left(\boldsymbol{E}_h(t),{c}_h(t)\right)\,=\,\frac{1}{2}\left(\boldsymbol{E}_h^T(t) \, \boldsymbol{Q}_h \, \boldsymbol{E}_h(t)\,+\,  {R}_h \, [{c}_h(t)]^2\right),
\end{eqnarray*}
where $\boldsymbol{Q}_h > \boldsymbol{0} \in \mathbb{R}^{3 \times 3}$ and ${R}_h > 0 \in \mathbb{R}$. The associated performance index is defined by
\begin{eqnarray}
J_h=\int_{0}^{\infty}U_h\left(\boldsymbol{E}_h(\zeta),{c}_h(\zeta)\right)\, d\zeta.
\label{Jh}
\end{eqnarray}

With the introduction of the tangential and bearing corrections, the trajectory-tracking optimization problem is reduced to minimizing the indices $J_x$, $J_y$, and $J_h$. It is important to point out that such a formulation avoids the explicit use of the full sixth-order nonlinear dynamics~\eqref{sysdyn}. 
The control signals $c_v$ and $c_h$ will be later calculated using a reinforcement learning process. These sub-control signals will be calculated in an online fashion based on location feedback measurements, such as GPS data, for instance. They are also employed implicitly into the computation of the thrust and rudder actuation forces $c_F$ and $c_R$.   
It is also worth mentioning that the term $\Angle{c_x + j c_y}$ may be omitted from the calculation of the control law $c_h(t)$. However, including it can speed up the convergence process.

\section{Optimal Control Solution}
\label{IRL_Bellman}
We now introduce the optimal control solution using integral forms of Bellman equations within a value iteration framework. The conditions of optimality are found using Bellman optimization principles~\cite{Lewis_2012}. The optimization problem finds the optimal control strategies by minimizing the performance indices $J_x$, $J_y,$ and $J_h$, with respect to the control signals $c_x$, $c_y,$ and $c_h$, respectively.

\subsection{Integral Bellman Equations}
The solving value structure for each segment of the optimization problem (i.e., searching for the optimal strategies  $c_x, \, c_y,$ and $c_h$) is motivated by the linear quadratic forms of the underlying cost functions. Hence, the solution for each segment is quadratic in the recent error records and sub-control signals such that
\begin{equation}
\label{model-free}  
J_i  \equiv V_i(\boldsymbol{E}_i,{c}_i)=\displaystyle \frac{1}{2} [\boldsymbol{E}_i ^T \,\,\,\,  {c}_i^T]\,\,\, \boldsymbol{H}_{i} \,\,\, 
\left[\begin{array}{l}
\boldsymbol{E}_i\\
{c}_i
\end{array}\right], 
\end{equation}
where $i \in \{x,y,h\},$ matrix $\boldsymbol{H}_i$, at each direction $x, \, y,$ and $h$, has a symmetric structure  $\boldsymbol{H}_i=\left[\begin{array} {ll}
\boldsymbol{H}_{\boldsymbol{E}_i\boldsymbol{E}_i}&\boldsymbol{H}_{\boldsymbol{E}_ic_i}\\
\boldsymbol{H}_{c_i \boldsymbol{E}_i}&\boldsymbol{H}_{c_ic_i}
\end{array}\right]>\boldsymbol{0}$.

Equating the performance indices~\eqref{model-free} to~\eqref{Jxy} and~\eqref{Jh}, yields the following integral Bellman equations
\begin{eqnarray}
V_i(\boldsymbol{E}_i(t) \, , \, {c}_i(t))=\int_{t}^{t+\Delta}U_i\left(\boldsymbol{E}_i(\zeta) \, , \, {c}_i(\zeta)\right)\, d\zeta \nonumber \\ \, + \,
V_i(\boldsymbol{E}_i(t+\Delta)\, , \,{c}_i(t+\Delta)),  i \in \{x,y,h\}.
\label{Bellman_Jxyh}
\end{eqnarray}
Applying the optimality conditions on these integral Bellman equations leads to the following optimal control strategies~\cite{Lewis_2012}:
\begin{eqnarray} \nonumber
{c}_i^*=\argmin_{{c}_i} \left(V_i(\boldsymbol{E}_i(t),{c}_i(t))\right),  i \in \{x,y,h\}. 
\label{IntBell}
\end{eqnarray}
Hence, each optimal strategy is evaluated such that $\boldsymbol{H}_{{c}_i{c}_i} {c}_i(t) + \boldsymbol{H}_{{c}_i\boldsymbol{E}_i} \boldsymbol{E}_i(t)=0$. The solution is a model-free optimal control strategy
\begin{equation}
{c}_i^*(t)=- \displaystyle \left[\boldsymbol{H}_{{c}_i{c}_i}^{-1} \, \boldsymbol{H}_{{c}_i\boldsymbol{E}_i}\right]  \boldsymbol{E}_i\,(t).
\label{optpol}
\end{equation}
Applying each optimal control strategy (i.e., $c_x^*,c_y^*,$ and $c_h^*$) into the respective integral Bellman equation~\eqref{Bellman_Jxyh} yields the following integral Bellman optimality equations %
\begin{eqnarray}
\label{BellOpt}  
V^*_i(\boldsymbol{E}_i(t),{c}^*_i(t))=\int_{t}^{t+\Delta}U_i\left(\boldsymbol{E}_i(\zeta),{c}^*_i(\zeta)\right)\, d\zeta+\nonumber\\ V^*_i(\boldsymbol{E}_i(t+\Delta),{c}^*_i(t+\Delta)),  i \in \{x,y,h\}.
\end{eqnarray}
The simultaneous solution of these integral Bellman optimality equations yields a trajectory-tracking for the sea vessel.

\subsection{Value Iteration Solution}
A value iteration algorithm is introduced to implement the online IRL solution for the Bellman optimality equations~\eqref{BellOpt}. This process does not require any initial admissible policies. It finds the optimal strategies using nondecreasing and upper-bounded sequence of solving value functions. The process is summarized in Algorithm~\ref{alg:alg1}. It is important to mention that the value iteration process employs partial knowledge about the dynamics of the sea vessel. It is proven to generally converge by generating a sequence of non-decreasing value functions that are upper-bounded by the optimal value~\cite{Abouheaf19}.

\begin{algorithm}
	\setstretch{1.2} 
	\caption{Online Value Iteration Mechanism}\label{alg:alg1}
	\begin{algorithmic}[1] 
		\Require
		\Statex Initial solving matrices ${\bf H}^1_i$, $i \in \{x,y,h\}$
		\Statex Initial tracking error vectors ${\bf E}^1_i$ and strategies $c^1_i$
		\Statex Error threshold $\gamma$
                \Statex Convergence time window $L$ 
		\Statex Maximum number of learning iterations $N_T$
		\Ensure
		\Statex Optimal solving matrices ${\bf H}^*_i, i \in\{x,y,h\}$
		\Statex
		\For{$\ell=1$ to $N_T$}
		\State Calculate the cost value $\int_{t}^{t+\Delta}U_i \, \left(\boldsymbol{E}^{\ell}_i(\zeta) \, , \, {c}^{\ell}_i(\zeta)\right)\, d\zeta$
		\State Find ${\bf E}^\ell_i(t+\Delta)$  and ${c}^\ell_i(t+\Delta)$ using \eqref{sysdyn} and \eqref{optpol}
		\State Determine $V^{\ell}_i(\boldsymbol{E}^\ell_i(t+\Delta) \, , \, {c}^{\ell}_i(t+\Delta))$ using~\eqref{model-free}
		\State Evaluate $V^{\ell+1}_i(\dots)$ using 	
		\begin{eqnarray*}	V^{\ell+1}_i(\dots)=\int_{t}^{t+\Delta}U_i \, \left(\boldsymbol{E}^\ell_i(\zeta) \, , \, {c}^{\ell}_i(\zeta)\right)\, d\zeta \nonumber \\ \, + \,
		V^{\ell}_i(\boldsymbol{E}_i(t+\Delta) \, , \, {c}^{\ell}_i(t+\Delta))
		\end{eqnarray*}
		\Comment{The policies $c_i$ and solving values $V_i$ will be later implemented using an adaptive critics scheme}	
		\If{$\ell > L$\AlgAnd $\norm{{\bf H}_i^{\ell+1-j}-{\bf H}_i^{\ell-j}} \le \gamma$, $\forall j \in \{0,1,\ldots,L\}$,}
		\State ${\bf H}_i^{*} \gets {\bf H}_i^{\ell+1}$
		\EndIf
		\EndFor
		\State \Return Optimal solving matrices ${\bf H}^*_i$
	\end{algorithmic}
\end{algorithm}

\section{Adaptive Critics Realization}
\label{actor-critic}
We adopt actor-critic structures in the form of neural network approximators employed by the IRL controller. The best strategies are approximated using actor networks, while the values of these strategies are approximated by means of critic networks~\cite{Widrow1973,Prokhorov1997,AbouheafCH2014}. Herein, each solving value function is approximated using a critic neural network defined by %
\begin{eqnarray}
\label{Crit}
	\hat V_i(\boldsymbol{E}_i,{\hat c_i})=\displaystyle \frac{1}{2} [\boldsymbol{E}_i^T \,\,\,\,  {\hat c_i}^T]\,\,\, \boldsymbol{W_i} \,\,\, 
	\left[\begin{array}{l}
		\boldsymbol{E}_i\\
		{\hat c_i}
	\end{array}\right],
\end{eqnarray}
where $\boldsymbol{W}_i$, $i \in \{x,y,h\}$ are the critic approximation weights for the solving value functions $\hat V_i$ (i.e., one for each of the three adaptive learning control loops).  
The structures of the critic networks are motivated by those of the  functions $\hat V_i$.
Similarly, the optimal strategies are approximated such that
\begin{eqnarray}
\label{pol}
\hat c_i= \boldsymbol{\Omega_i} \boldsymbol{E}_i,
\end{eqnarray}
where $\boldsymbol{\Omega}_i,i\in\{x,y,h\}$ are network approximation weights.

The adaption process of the adaptive critics weights employs a gradient descent approach. The target values which are used to tune the different critic weights are given by
\begingroup\makeatletter\def\f@size{9.2}\check@mathfonts
\def\maketag@@@#1{\hbox{\m@th\large\normalfont#1}}
\begin{equation}
\small
\tilde V_i=\int_{t}^{t+\Delta}U_i\left(\boldsymbol{E}_i(\zeta),\hat{c}_i(\zeta)\right)\, d\zeta +
\hat V_i(\boldsymbol{E}_i(t+\Delta),\hat {c}_i(t+\Delta)) .
\label{Valt}
\end{equation}
%
They express the desired value functions of the approximations $\hat V_i$, $i\in\{x,y,h\}$. Similarly, the desired values of the approximated optimal strategies $\hat c_i$, that are used to tune the actor weights, are defined by
\begin{eqnarray}\tilde c_i=-\left[\boldsymbol{W}^{-1}_{{\hat c_i}  {\hat c_i}} \boldsymbol{W}_{{\hat c_i}\boldsymbol{E}_i}\right] \boldsymbol{E}_i, ~~ i\in\{x,y,h\}.
\label{Polt}
\end{eqnarray}
The approximation error of each critic network is defined by
$\varepsilon^{Critic}_i=\frac{1}{2}\left(\hat V_i(\boldsymbol{E}_i,{\hat c}_i)-\tilde V_i\right)^2,$ while the approximation error of each actor network is given by $\varepsilon^{Actor}_i=\frac{1}{2}\left(\hat c_i-\tilde c_i\right)^2$.
The neural network weights are tuned through a gradient descent approach. It yields the following update rules for the critic approximation weights
$\boldsymbol{W}_i$, $i\in\{x,y,h\}$:
\begin{equation} \label{NNval1}
\small
\boldsymbol{W}_i^{\ell+1} =\boldsymbol{W}_i^{\ell} -\eta_c \left(\left(\hat V_i(\boldsymbol{E}_i,{\hat c_i}) - \tilde V_i\right) \left[\begin{array}{l}
\boldsymbol{E}_i\\
{\hat c}_i
\end{array}\right] [\boldsymbol{E}_i^T \,  {\hat c}_i^T]\right)^{\ell},
\end{equation}
%
where $0 < \eta_c < 1$ is a critic-network learning rate.
Similarly, the update laws for the actor weights are
\begin{equation} \label{NNpol1} 
\small
\boldsymbol{\Omega}_i^{\ell+1} =\boldsymbol{\Omega}_i^{\ell} -\eta_a \left(\left({\hat c}_i -  \boldsymbol{\tilde c}_i  \right) \boldsymbol{E}^{T}_i \right)^{\ell},
\end{equation}
%
where  $0 < \eta_a < 1$ is an actor-network learning rate. The actor and critic weights are tuned online using a value iteration process as detailed out in Algorithm~\ref{alg:alg2}.

\begin{algorithm}
	\setstretch{1.2} 
	\caption{Adaptive Critics Implementation}\label{alg:alg2}
	\begin{algorithmic}[1] 
		\Require
		\Statex Initial neural network weights ${\bf W}^1_i$ and ${\bf \Omega}^1_i, i \in \{x,y,h\}$
		\Statex Initial tracking error vectors ${\bf E}^1_i$ and strategies $\hat c^1_i$
		\Statex Error threshold $\gamma$
                \Statex Convergence time window $L$ 
		\Statex Maximum number of learning iterations $N_T$
		\Ensure
		\Statex Tuned neural network weights ${\bf W}^*_i$ and ${\bf \Omega}^*_i, i \in \{x,y,h\}$
		\Statex
		\For{$\ell=1$ to $N_T$}
		\State Calculate the cost value $\int_{t}^{t+\Delta}U_i \, \left(\boldsymbol{E}^{\ell}_i(\zeta) \, , \, {\hat c}^{\ell}_i(\zeta)\right)\, d\zeta$
		\State Find ${\bf E}^\ell_i(t+\Delta)$  and $\hat c^\ell_i(t+\Delta)$ using~\eqref{sysdyn} and~\eqref{pol}
		\State Compute $\hat V^{\ell}_i(\boldsymbol{E}^\ell_i(t+\Delta) \, , \, \hat c^{\ell}_i(t+\Delta))$ using~\eqref{Crit}
		\State Determine $\tilde V_i$ and $\tilde c_i$ using~\eqref{Valt} and~\eqref{Polt}, respectively	
		\State Update the critic and actor weights using~\eqref{NNval1} and~\eqref{NNpol1}, respectively
		\If{$\ell > L$\AlgAnd $\norm{{\bf W}_i^{\ell+1-j}-{\bf W}_i^{\ell-j}}\le \gamma$, $\forall j \in \{0,1,\ldots,L\}$,}
		\State ${\bf W}_i^{*} \gets {\bf W}_i^{\ell+1}$
		\EndIf
                \If{$\ell > L$\AlgAnd $\norm{{\bf \Omega}_i^{\ell+1-j}-{\bf \Omega}_i^{\ell-j}}\le \gamma$, $\forall j \in \{0,1,\ldots,L\}$,}
		\State ${\bf \Omega}_i^{*} \gets {\bf \Omega}_i^{\ell+1}$
		\EndIf
		\EndFor
		\State \Return Tuned weights ${\bf W}^*_i$ and ${\bf \Omega}^*_i$
	\end{algorithmic}
\end{algorithm}

\section{Simulations and Results}
\label{sim}
The usefulness of the proposed IRL-based adaptive learning approach is verified using two simulation scenarios. In the first scenario, the sea vessel tracks a linear reference trajectory with a constant speed. In the second scenario, the vessel follows a dynamic reference trajectory with varying linear and angular velocities. 
We use the same dynamic parameters of the vessel as in~\cite{Ding2011}:
$m_{11}=\SI{19}{\kilogram}$,
$d_{11}= \SI{4}{\kilogram/\s}$,
$m_{22}=\SI{35.2}{\kilogram}$,
$d_{22}=\SI{1}{\kilogram/\s}$,
$m_{33}=\SI{4.2}{\kilogram . \meter^2}$, and
$d_{33}=\SI{10}{\kilogram . \m^2/ \s}$.
The parameters of the learning environment are chosen as
$\Delta=\SI{0.1}{\s}$, $\eta_c=0.005$, $\eta_a=0.1$.
The weighting matrices are set to
$\boldsymbol{Q}_i=\boldsymbol{I}_{3\times3}$ and ${R}_i =1$, $i\in\{x,y,h\}$, where $\boldsymbol{I}$ is the identity matrix.

\subsection{Scenario~1: Linear Trajectory with a Constant Speed}
In this scenario, the vessel is set to follow a trajectory defined by
$v^{\mathrm{ref}}(t)=\SI{9}{\meter / \s}$ and $\psi^{\mathrm{ref}} (t)=\SI{0.5}{\radian}$, $\forall t \geq 0$,
with initial conditions
$( x^{\text{ref}}(0)  ,  y^{\text{ref}}(0) )=(40,60)$ and $( x(0)  ,  y(0) )=(-100,-100)$.
Fig.~\ref{fig:case1critic} reveals the convergence of the variations in the critic weights. The values of the thrust and rudder control forces are shown in Fig.~\ref{fig:case1_control}. They are synchronized with the  critic weights adaptation results. The vessel starts off by picking up speed till it latches to the reference trajectory in less than \SI{50}{\s}. This is clearly seen in Figs.~\ref{fig:case1velocity} and~\ref{fig:case1orient}.

\begin{figure}
	\vspace{-15pt}
	\hspace{-25pt}\includegraphics[width=10.5cm, height=9.5cm]{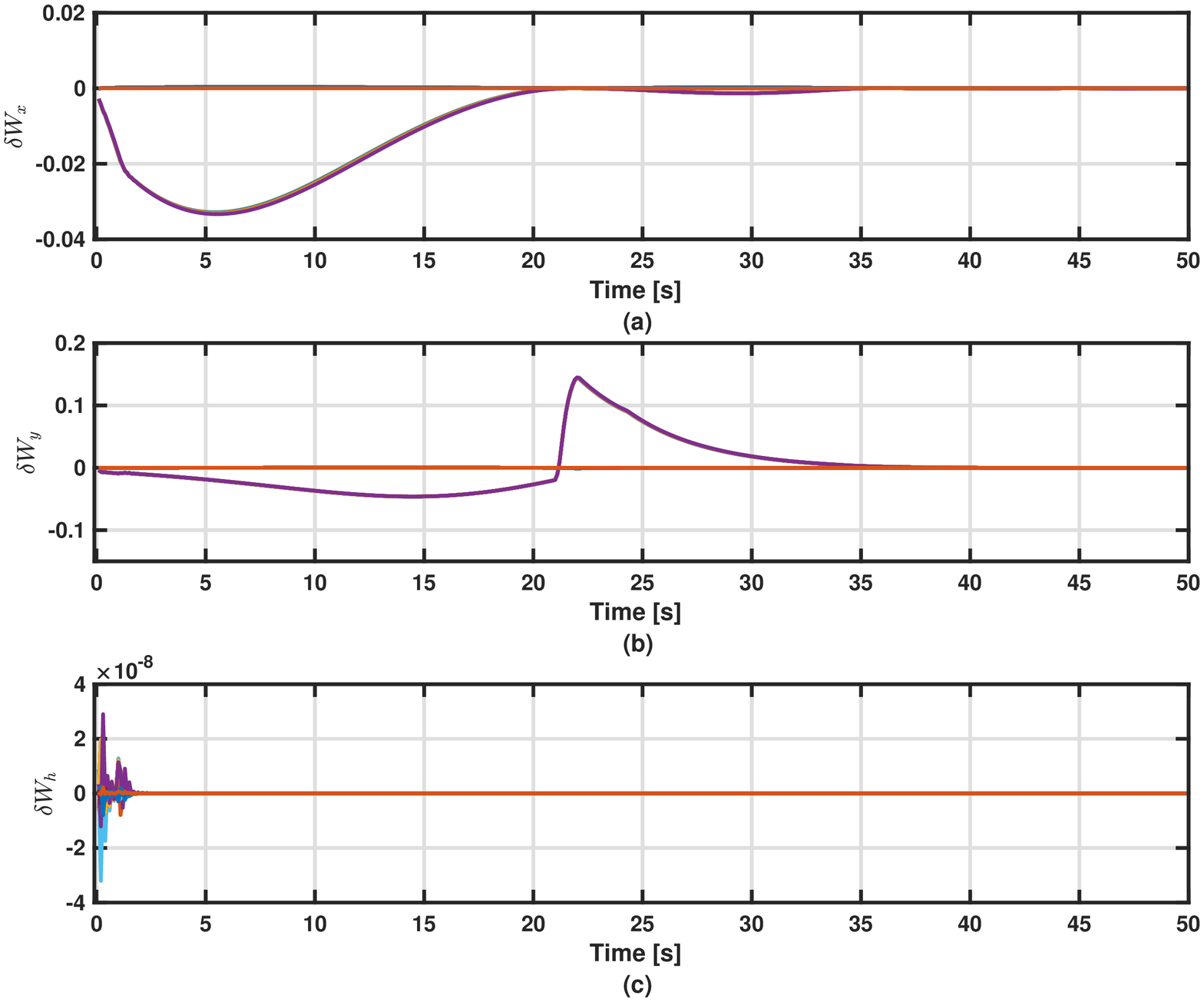}
	 \vspace{-45pt}
	\caption{Scenario~1: variations in critic weights, (a)~$\delta W_x$, (b)~$\delta W_y$, and (c)~$\delta W_h$.}
	\label{fig:case1critic}
\end{figure}
\begin{figure}[htb]
	\centering
	\includegraphics[width=1\linewidth]{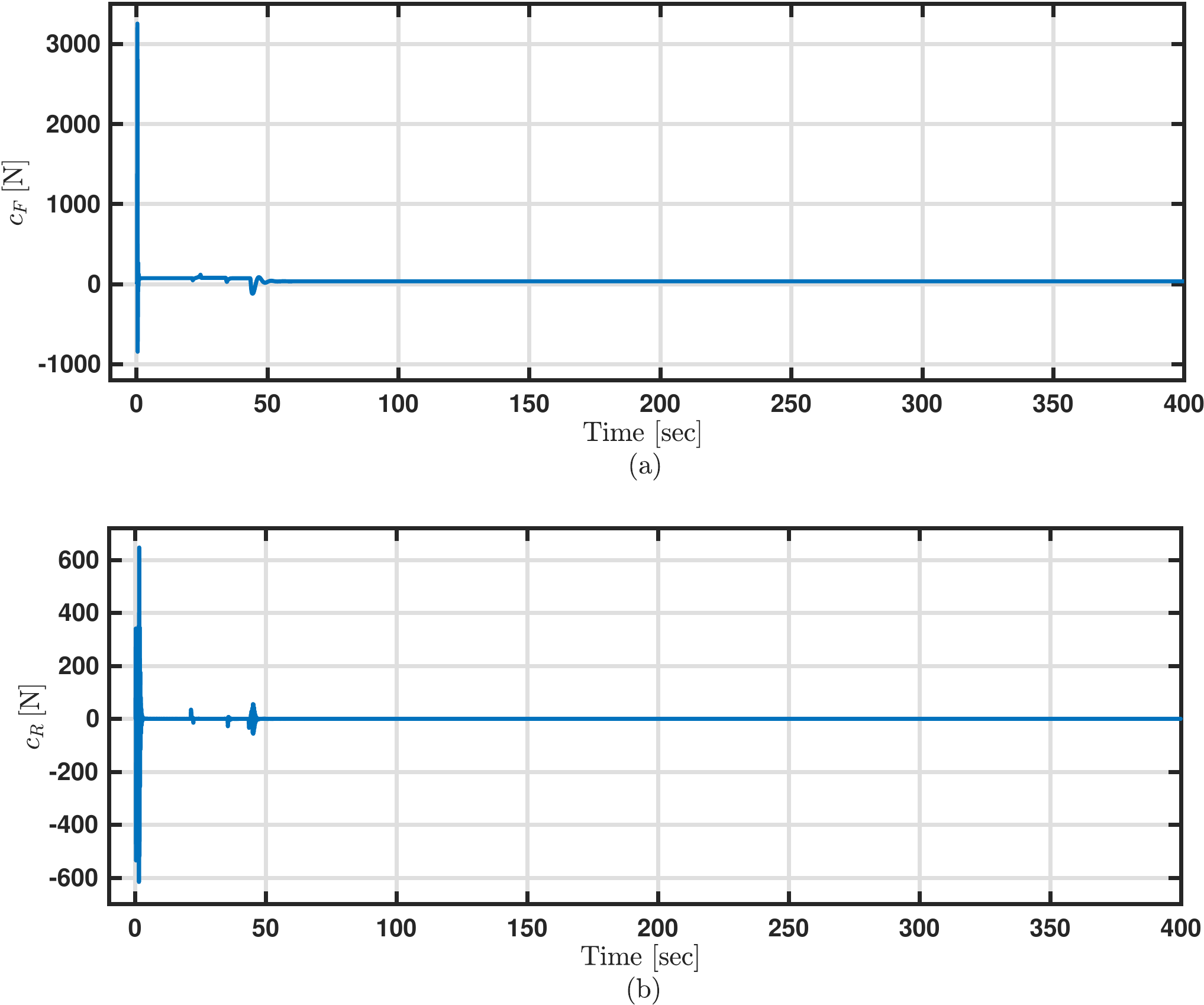}
	\vspace{-20pt}
	\caption{Scenario~1: (a)~Thrust control force $c_F$,  (b)~Rudder control force $c_R$.}
	\label{fig:case1_control}
\end{figure}
\begin{figure}[htb]
	\centering
	\includegraphics[width=0.98\linewidth]{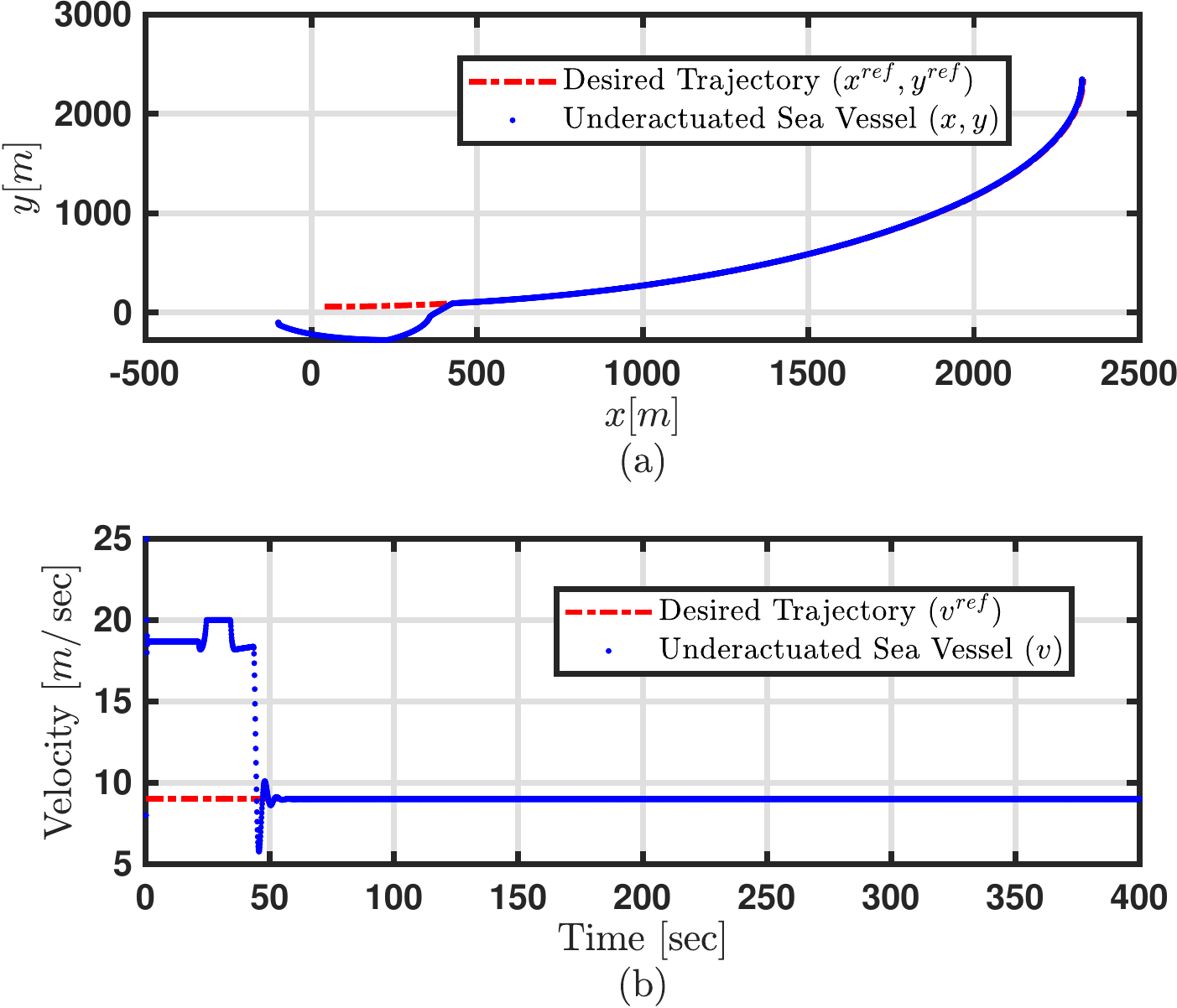}
	 \vspace{-10pt}
	\caption{Scenario~1: (a)~Position phase plan, (b)~Linear velocity.}
	\label{fig:case1velocity}
\end{figure}
\begin{figure}[htb]
	\centering
	\includegraphics[width=0.98\linewidth]{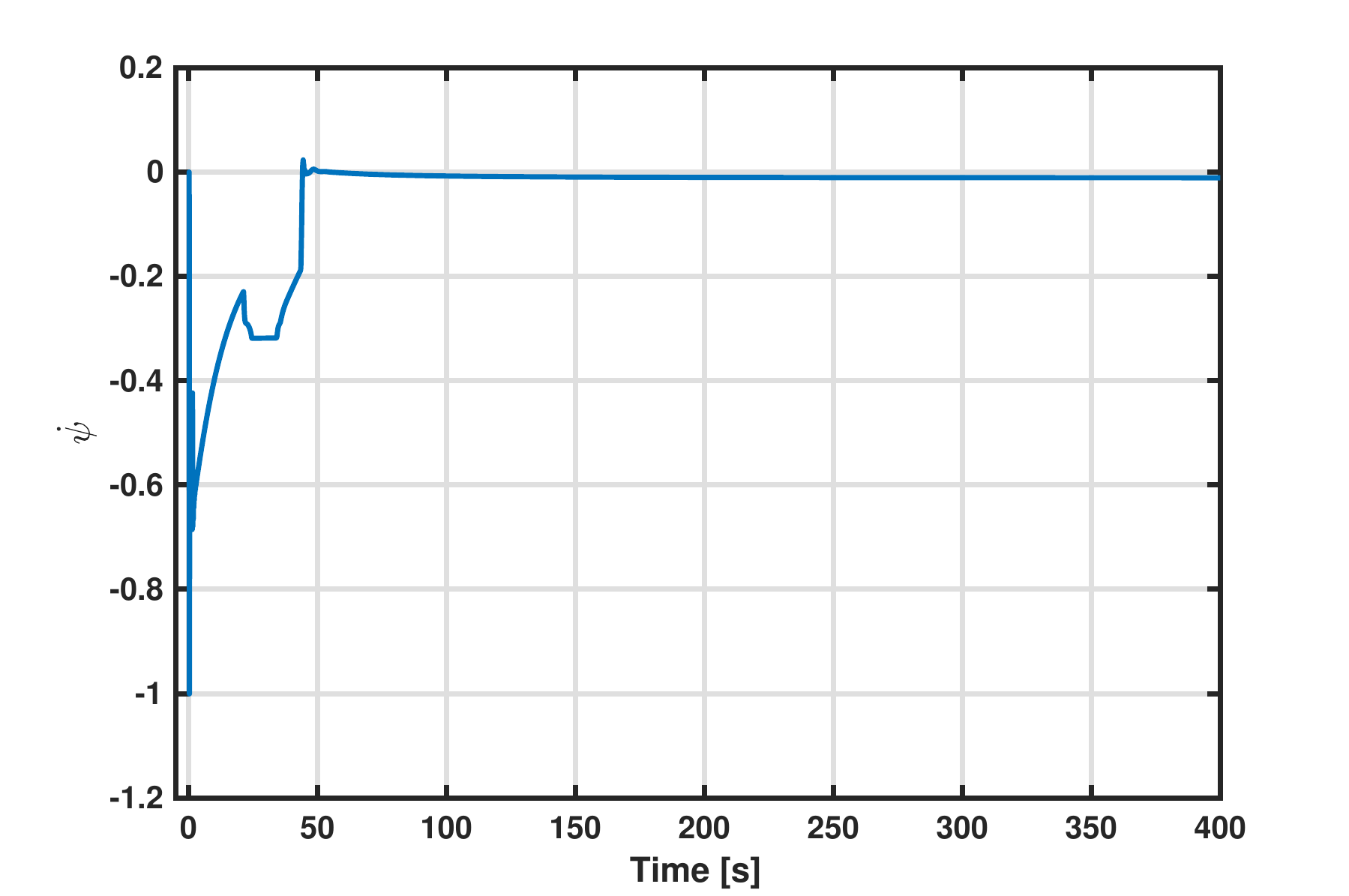}
	 \vspace{-10pt}
	\caption{Scenario~1: Angular velocity.}
	\label{fig:case1orient}
\end{figure}

\subsection{Scenario~2: Reference Trajectory with Time-Varying Linear and Angular Velocities}
In this experiment, the vessel is set to track a challenging trajectory defined by
$v^{\mathrm{ref}}\,(t)=9+0.2 \, n(t) \, (15\, \cos\, (5 \, \pi \Delta \, t) +5\, \sin\, (5 \, \pi \Delta \, t))$ and $\psi^{\mathrm{ref}}\,(t)=0.5\, n(t) \, \cos\,(20\,\pi\,\Delta \,t)$.
The parameter $n(t)$ is a random variable drawn from a Gaussian distribution $\sim \mathcal{N}(0,1)$.
The initial conditions are taken as
$( x^{\text{ref}}(0)  ,  y^{\text{ref}}(0) )=(40,60)$ and
$( x(0)  ,  y(0) )=(30,60)$.
The thrust and rudder forces are shown in Fig.~\ref{fig:case2_control}. The adaptive learning process enabled the vessel to track the rapidly varying sinusoidal trajectory, as illustrated in Figs.~\ref{fig:case2velocity} and~\ref{fig:case2orient}.
The figures unveil the ability of the adaptive learning process to adjust to the high maneuverability enforced by the reference trajectory.

\begin{figure}[htb]
	\centering
	\includegraphics[width=1\linewidth]{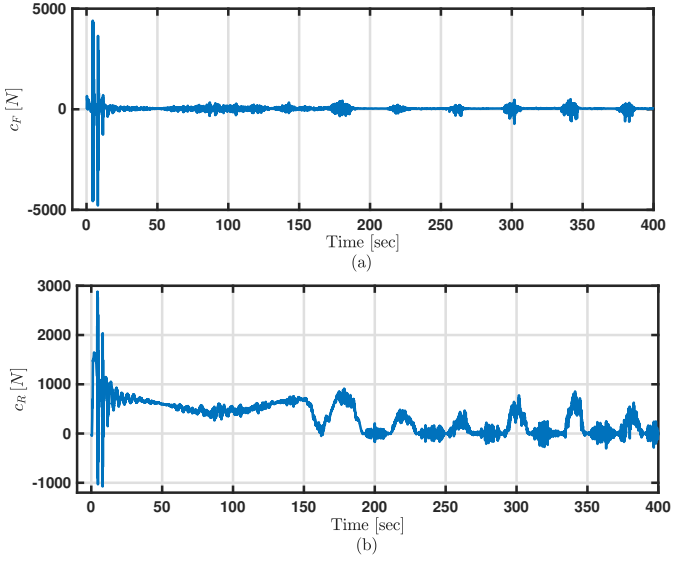}
	 \vspace{-10pt}
	\caption{Scenario~2: (a)~Thrust control force $c_F$,  (b)~Rudder control force $c_R$.}
	\label{fig:case2_control}
\end{figure}
\begin{figure}[htb]
	\centering
	\includegraphics[width=1\linewidth]{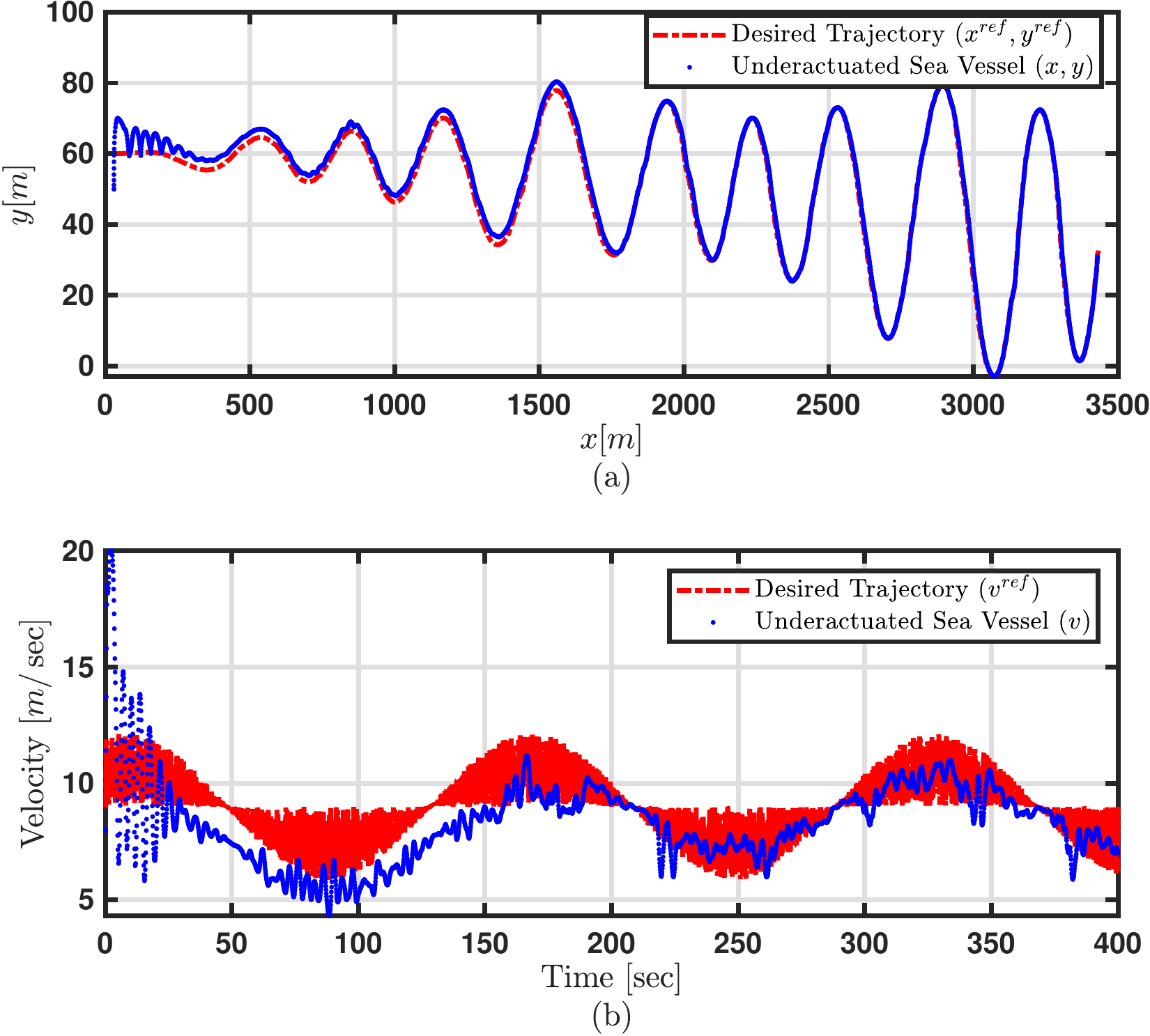}
	 \vspace{-10pt}
	\caption{Scenario~2: (a)~Position phase plan, (b)~Linear velocity.}
	\label{fig:case2velocity}
\end{figure}
\begin{figure}[htb]
	\centering
	\includegraphics[width=1\linewidth]{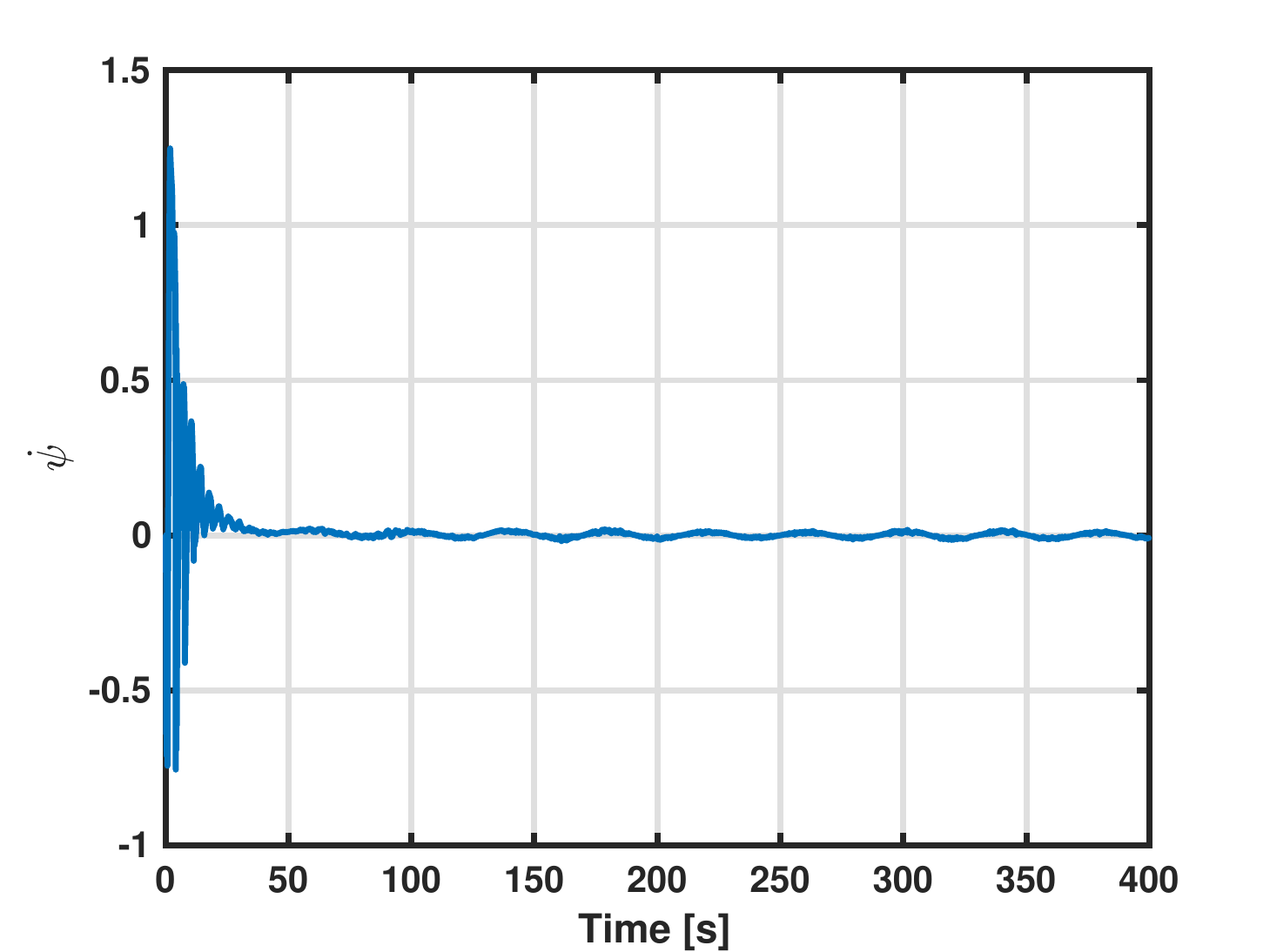}
	 \vspace{-10pt}
	\caption{Scenario~2: Angular velocity.}
	\label{fig:case2orient}
\end{figure}

\section{Conclusion}
\label{conc}
The paper introduces an online integral reinforcement learning mechanism to control an underactuated sea vessel. The solution employs a value iteration process that uses an integral form of Bellman equation. This approach does not employ any traditional error dynamics equations which typically result in hard-to-implement control policies. It rather makes use of measurements relevant to the position of the vessel to make optimal control decisions. Further, it relaxes the dependence of the solution on recognizing the complete dynamical model of the vessel by suggesting intermediate model-free control strategies. The thrust and rudder actuation forces of the vessel are set to be explicit functions of such strategies. The adaptive critics are then used to implement the integral reinforcement learning solution by approximating the underlying optimal strategies and their associated values. The performance of the adaptive learning process is validated using two test cases.

\bibliographystyle{IEEEtran}
\bibliography{Bib/mybibliography}

\end{document}